\documentclass[10pt,conference,comsoc,final]{IEEEtran}

\usepackage[pdftex]{graphicx}
\usepackage{amsmath}
\usepackage[cmintegrals]{newtxmath}
\usepackage{algpseudocode}
\usepackage{algorithm}
\usepackage{array}
\usepackage{url}
\usepackage{lipsum}
\usepackage{booktabs} %
\usepackage[caption=false,font=footnotesize]{subfig} %
\usepackage{siunitx} %
\usepackage{color,soul} %
\usepackage{siunitx} %

\makeatletter
\def\ps@IEEEtitlepagestyle{%
   \def\@oddfoot{\mycopyrightnotice}%
  \def\@evenfoot{}%
}
\def\mycopyrightnotice{%
  {\footnotesize\hfill Published in 2021 IEEE Wireless Communications and Networking Conference (WCNC), DOI: 10.1109/WCNC49053.2021.9417575. \copyright~2021 IEEE\hfill}%
  \gdef\mycopyrightnotice{}%
}
\makeatother

\makeatletter
\newcommand\fs@betterruled{%
  \def\@fs@cfont{\bfseries}\let\@fs@capt\floatc@ruled
  \def\@fs@pre{\vspace*{5pt}\hrule height.8pt depth0pt \kern2pt}%
  \def\@fs@post{\kern2pt\hrule\relax}%
  \def\@fs@mid{\kern2pt\hrule\kern2pt}%
  \let\@fs@iftopcapt\iftrue}
\floatstyle{betterruled}
\restylefloat{algorithm}
\makeatother

\def\BibTeX{{\rm B\kern-.05em{\sc i\kern-.025em b}\kern-.08em T\kern-.1667em\lower.7ex\hbox{E}\kern-.125emX}}

\begin{document}
\newcommand{\myScaling}{0.42}

\title{Contention Window Optimization in IEEE 802.11ax Networks with Deep Reinforcement Learning}

\author{\IEEEauthorblockN{Witold Wydma\'nski}
\IEEEauthorblockA{Faculty of Computer Science, Electronics\\ and Telecommunications\\ AGH University of Science and Technology\\
Krakow, Poland \\ wwydmanski@gmail.com}
\and
\IEEEauthorblockN{Szymon Szott}
\IEEEauthorblockA{Faculty of Computer Science, Electronics\\ and Telecommunications \\ AGH University of Science and Technology\\
Krakow, Poland \\
szott@agh.edu.pl}
}

\noindent\parbox[t]{\textwidth}{
{\Large Please cite this paper as: 

W. Wydmański and S. Szott, ``Contention Window Optimization in IEEE 802.11ax Networks with Deep Reinforcement Learning,'' 2021 IEEE Wireless Communications and Networking Conference (WCNC), 2021. \url{https://doi.org/10.1109/WCNC49053.2021.9417575}
}
}

\vspace{2cm}
\begin{verbatim}
@INPROCEEDINGS{wydmanski2021contention,
  author={Wydmański, Witold and Szott, Szymon},
  booktitle={2021 IEEE Wireless Communications and Networking Conference (WCNC)}, 
  title={{Contention Window Optimization in IEEE 802.11ax Networks with 
  Deep Reinforcement Learning}}, 
  year={2021},
  volume={},
  number={},
  pages={1--6},
  doi={10.1109/WCNC49053.2021.9417575}}
\end{verbatim}
\clearpage

\maketitle

\begin{abstract}
The proper setting of contention window (CW) values has a significant impact on the efficiency of Wi-Fi networks. Unfortunately, the standard method used by 802.11 networks is not scalable enough to maintain stable throughput for an increasing number of stations, yet it remains the default method of channel access for 802.11ax single-user transmissions. Therefore, we propose a new method of CW control, which leverages deep reinforcement learning (DRL) principles to learn the correct settings under different network conditions. Our method, called centralized contention window optimization with DRL (CCOD), supports two trainable control algorithms: deep Q-network (DQN) and deep deterministic policy gradient (DDPG). We demonstrate through simulations that it offers efficiency close to optimal (even in dynamic topologies) while keeping computational cost low. 
\end{abstract}

\section{Introduction}

\IEEEPARstart{T}{he} upcoming IEEE 802.11ax amendment, scheduled for release in 2020, has a goal of increasing Wi-Fi network efficiency \cite{Khorov2019}. However, to ensure backward compatibility, one efficiency-related aspect remains unchanged in 802.11ax: the basic channel access method \cite{Bellalta2019}. This method is an implementation of carrier-sense multiple access with collision avoidance (CSMA/CA), wherein each station ``backs off'', i.e., waits a certain number of time slots, before accessing the channel. This number is chosen at random from the interval 0 to CW (the contention window). To reduce the probability of multiple stations selecting the same random number, CW is doubled after each collision. IEEE 802.11 defines static CW minimum and maximum values and this approach, while being robust to network changes and requiring few computations, can lead to inefficient operation, especially in dense networks \cite{Gallo2018}.  

CW optimization has a direct impact on network performance %
and as such has been the subject of multiple research analyses. Example optimization approaches include using control theory \cite{Serrano2013} and monitoring the number of active users \cite{Karaca2017}. 
With the proliferation of network devices with high computational capabilities, CW optimization can now be analyzed using 
machine learning (ML) methods \cite{Wilhelmi2019,Sandholm2019,30years}.
The choice of available ML algorithms is limited by the nature of the problem. Analytical models, such as \cite{Bianchi2000}, can provide optimal CW values but only under certain assumptions and quasi-static settings. 
The most popular approach, supervised learning, depends on minimizing the difference between the inferred result and a specified, optimal solution. Since the assumptions for analytical models are rarely encountered in real networks, we cannot depend on them for training the model -- if we could, the model would be unnecessary. Another common approach is gradient-based optimization. If we were able to calculate a derivative of the function of network throughput with respect to CW, we could effortlessly find an optimal value. Unfortunately, this is not the case.
However, we can still use a wide range of so called \textit{derivative-free optimization algorithms}.
In particular, reinforcement learning (RL) is a branch of ML well-suited to the problem of improving the performance of wireless networks because it deals with intelligent software agents (network nodes) taking actions (e.g., optimizing parameters) in an environment (wireless radio) to maximize a reward (e.g., throughput) \cite{Zhang2019}. 
RL is an example of model-free policy optimization, offering better generalization capabilities than conventional, model-based optimization approaches such as control theory\footnote{This means that while RL algorithms try to directly learn an optimal policy without learning the model of the environment, model-based approaches need to make assumptions about the model's next state before choosing an action.}.

In this paper, we describe \emph{centralized contention window optimization with DRL} (CCOD), our proposed method of applying DRL to the task of optimizing saturation throughput of 802.11 networks by correctly predicting CW values. While CCOD is universally applicable to any 802.11 network, we exhibit its operation under 802.11ax using two DRL methods: deep Q-network (DQN) \cite{Mnih2015} and deep deterministic policy gradient (DDPG) \cite{DPG}. The former is considered a showcase DRL algorithm, while the latter is a more advanced method, able to directly learn the optimal policy, which we expect will lead to increased network performance, especially in dense scenarios.  
Additionally, we demonstrate how we applied time series analysis to the recurrent neural networks of both DRL methods. Finally, we provide the complete source code so that the work can serve as a stepping stone for further development of DRL-based methods in 802.11 networks\footnote{\url{https://github.com/wwydmanski/RLinWiFi}}.%

\section{Related Work}
A recent example of applying RL to wireless local area networks include a jamming countermeasure \cite{Yao2019} and an ML-enabling architecture \cite{Wilhelmi2019}. 
RL performance can be further improved by using deep artificial neural networks with their potential for interpolation and superior scalability.
Recent examples of using deep RL (DRL) in wireless networks include: a general adaptable medium access control (MAC) protocol \cite{Yu2019}, a MAC protocol for underwater acoustic networks \cite{under-the-sea}, a radio resource scheduling protocol \cite{altam2020learn}, and a method for tuning reconfigurable antennas \cite{Huang2020}. 
The authors of \cite{Ali2019} also claim to use DRL in the area of CW optimization. However, a careful reading reveals that they use Q-learning (a typical RL method) but without the neural network (deep) component. 
Thus, we conclude that DRL has not yet been successfully applied to study IEEE 802.11 CW optimization.

\section{Applying DRL to Wi-Fi Networks}
RL is based on the idea that instead of using an analytical approach to solve a problem, we can use a self-learning agent, which is able to interact with the environment via a series of \textit{actions}.
Each of these actions bring the agent closer to (or further from) its goal -- for which it will gain a \textit{reward}.
Through a training process, the agent enhances its \textit{decision-making policy} %
until it learns %
the best possible decision in every state of the environment that the agent can visit.
In DRL, the agent's policy is determined by a deep neural network which requires training.
We consider two DRL methods differing on their action space: discrete (DQN) and continuous (DDPG).

DQN is based on Q-learning \cite{Zhang2019}, which attempts to predict an expected reward for each action, making it an example of a \textit{value-based} method. 
DQN's additional deep neural network allows for more efficient extrapolation of rewards for yet unseen states as compared to basic Q-learning.

Conversely, DDPG is an example of a \textit{policy-based} method, because it tries to learn the optimal policy directly. 
Additionally, it can produce unbounded \textit{continuous} output meaning that it can recognize that the action space is an ordered set (as in the case of CW optimization)\footnote{Discrete algorithms, like DQN, consider all possible actions as abstract alternatives.}.
DDPG comprises two neural networks: an \textit{actor} and a \textit{critic}. The actor makes decisions based on the environment state, while the critic is a DQN-like neural network that tries to learn the expected reward for the actor's actions.

To apply DRL principles to Wi-Fi networks, we propose the CCOD %
method, which comprises an agent, the environment states, the available actions, and the received rewards. In summary, the CCOD \textit{agent} is a module which observes the \textit{state} of the Wi-Fi network, selects appropriate CW values (from the available \textit{actions}) in order to maximize network performance (the \textit{reward}).

\subsection{Formal Definition}
This setup requires framing the optimization problem as a Markov decision process (MDP) which consists (among others) of defining an agent, states, actions, and rewards. 
This problem, however, can be better described using an MDP generalization -- a partially observable Markov decision process (POMDP), which does not assume that we can observe the environment's state perfectly.
The POMDP is formally a $(S, A, T, R, \Omega, O, \gamma)$ tuple, in which:
\begin{itemize}
  \item $S$ is the set of states,
  \item $A$ is the set of actions,
  \item $T$ is the set of probabilities regarding transition between states,
  \item $R: S \times A \mapsto \mathcal{R}$ is the reward function,
  \item $\Omega$ is the set of observations,
  \item $O$ is the set of observation probabilities,
  \item $\gamma$ is the discount coefficient.
\end{itemize}
We describe each element of the tuple below.

The \textit{agent} is located at the access point (AP), because the AP has a global view of the network, it can control its associated stations in a centralized manner through beacon frames, and it can handle the computational requirements of DRL's agent-based inference. Furthermore, a CCOD AP can potentially exchange information with other APs and become part of an SDN-based multi-agent Wi-Fi architecture \cite{Gallo2018}. 

The \textit{state} $s \in S$ of the environment is the exact status of all of the devices currently connected to the network. Because of the nature of the problem it is impossible to gather this information. Thus, the problem is presented as a POMDP instead of a MDP.

The \textit{action} $a \in A$ of the agent corresponds directly to setting the new CW value. Recall that CW defines the upper bound when selecting the random number of backoff slots (9~\si{\micro\second} each) to count down in 802.11's channel access function. We explore two algorithms with different types of output -- discrete and continuous. Discrete output results in an integer $a$ between 0 and 6. Continuous output produces a real number $ a \in [0, 6] $. The output is then used to update $CW$ as follows:
\begin{equation} \label{eq:CW}
    CW = \left\lfloor2^{a+4}\right\rfloor-1
\end{equation}
The range has been chosen so that $CW$ fits into the original span of 802.11 values: from 15 to 1023. Therefore, the set of all possible actions $A$ can be defined as $A = [0, 6]$.
Each decision $a$ regarding the CW value taken by the agent in state $s$ causes the environment to transition to state $s' \in S$ with probability $T\left(s'|s,a\right)$. 

We use network throughput (the number of successfully delivered bits per second) as the \textit{reward} $r \in [0, 1]$ in CCOD. This is indicative of the current network performance and can be observed at the AP. Since rewards in DRL should be a real number between 0 and 1, we divide observed bits per second by the expected maximum throughput.

The \textit{observation} $o \in \Omega$ characterizing the environment has to give the maximum possible insight into the current status of the network. We define each observation as the current collision probability $p_{col}$ (the transmission failure probability) observed in the Wi-Fi network calculated based on the number of transmitted frames $N_{t}$ and correctly received frames $N_{r}$:
\begin{equation}
    p_{col}=\frac{N_{t}-N_{r}}{N_{t}}.
\end{equation}
The $p_{col}$ measurements are done within predefined \textit{interaction periods} and reflect the performance of the currently selected CW value. In practice, $p_{col}$ is not immediately available to the agent, but since the AP takes part in all frame transmissions (as sender or recipient), the agent knows the AP's $N_{t}$ and requires only obtaining $N_{t}$ from each station, which can be piggybacked onto data frames. Regarding $N_{r}$, it is known at the AP based on the number of sent or received acknowledgement frames. 
Note that in the performance analysis we consider only uplink (i.e., station-initiated) transmissions (Fig.~\ref{fig:topology}).

Finally, to determine the importance of future rewards in comparison to immediate rewards, the discount coefficient $\gamma$ is used. This hyperparameter is adjustable by the user.

\subsection{CCOD as a CW Control Method}
CCOD operates in three phases. 
In the first, \textit{pre-learning phase}, the Wi-Fi network is controlled by legacy 802.11. This serves as a warm-up for CCOD's DRL algorithms. 
Afterwards, in the \textit{learning phase}, the agent undertakes decisions regarding the CW value following %
Algorithm~\ref{alg: cw_optim}.

The preprocessing in the algorithm consists of calculating the mean and standard deviation of the history of recently observed collision probabilities $H(p_{col})$ of length $h$ using a moving window of a fixed size and stride. 
This operation changes the data's shape from one- to two- dimensional (each step of the moving window yields two data points). This collection can then be interpreted as a time series, which means it can be analysed by a recurrent neural network. Their design allows for a more in-depth understanding of both the immediate and indirect relations between agent actions and network congestion compared to a one-dimensional analysis with a dense neural network.

To enable exploration, each action is modified by a noise factor, which decays over the course of the learning phase. For DQN, noise is the probability of overriding the agent's action with a random action. For DDPG, noise is sampled from a Gaussian distribution and added to the decision of the agent. 
We do this to mitigate an \textit{exploration-exploitation} issue -- finding a balance between looking for new experiences (exploration) and maximizing the reward via the best known action (exploitation).

The final, \textit{operational phase} starts after completing training, which is determined by a user-set time limit. Now the noise factor is equal to zero, and the agent will always choose the best possible action.
The agent is considered to be fully trained and will no longer receive any updates, so \textit{rewards} are no longer needed. The agent is ready to be deployed in a network.
If the need arises, further training of the agent can be done in a distributed manner \cite{gupta2018distributed} -- a copy of the agent can be retrained in a controlled simulation, after which the weights of both agents can be averaged.
\setlength{\textfloatsep}{10pt}
\begin{algorithm}
\caption{CW optimization using CCOD}
\label{alg: cw_optim}
\begin{algorithmic}[1]
\item[]
\State {Initialize observations buffer $B_{obs}$ with zeroes}
\State {Get the agent weights $\theta$}
\State {Get the agent action function $A_{\theta}$}
\State {Define $N_r$, $N_t$ as number of received, transmitted frames}
\State {Define $\Delta t$ as interaction period}
\State {Define $train$ as a flag whether the algorithm is in training}
\State {Define $B$ as the replay buffer}
\\
\State {$last\_update \gets current\_time$ }
\State {$CW \gets 15$}
\State {$s \gets $ vector of zeroes}
\\
\For{t = 1, ..., $\infty$}
    \State {$N_t \gets$ number of transmitted frames}
    \State {$B_{obs}.$push$(N_r/N_t)$}
    \If {$last\_update + \Delta t \leq current\_time$ }
        \State {$obs \gets$ preprocess$(B_{obs})$}
        \State {$a \gets A_{\theta}(obs)$}
        \State {$CW \gets 2^{a+4}$-1}
        \If {$train$}
            \State {$Thr \gets N_r/\Delta t$  }
            \State {$r \gets$ normalize$(Thr)$}
            \State {$B.$push$((obs,\ a,\ r,\ s))$}
            \State {$s \gets obs$}
            \State {$b \gets$ sampled minibatch from $B$}
            \State {Perform optimization on $\theta$ basing on $b$}
        \EndIf
    \EndIf

\EndFor
\end{algorithmic}
\end{algorithm}
The application of DRL algorithms also requires configuring certain key parameters. First, the performance of RL algorithms depends on their \textit{reward discounts} $\gamma$, which correspond to the importance of long term rewards over immediate ones. Second, the introduction of deep learning into RL algorithms creates an impediment in the form of many new hyperparameters so each neural network requires configuring a \textit{learning rate} %
as an update coefficient. Third, since the learning is done by mini-batch stochastic gradient descent, the correct choice of \textit{batch size} is also critical. Finally, both algorithms use a \textit{replay buffer} $B$, which records every interaction between the agent and the environment (up to a size limit), and serves as a base for mini-batch sampling. The replay buffer stores the interactions in the form of (state, action, reward, next state) tuples.

Additionally, both algorithms apply separation between the \textit{local} and \textit{target} neural networks for smoothing out the reward noise -- the actions are decided by the \textit{local} network, but the learning algorithm uses predictions from the \textit{target} network. Then, the weights of the \textit{target} network are set to a weighted average of the two based on the $\tau$ parameter:
$w_{target}=\tau\cdot w_{local} + \left(1-\tau\right)\cdot w_{target}$.

\section{Simulation Model}

We implemented CCOD in ns3-gym \cite{ns3gym}, which is a framework for connecting ns-3 %
(a network simulator) with OpenAI Gym %
(a tool for DRL analysis). 
The neural networks of DDPG and DQN were implemented in Pytorch %
and Tensorflow, %
respectively, to demonstrate independence of agent implementation.

\setlength{\textfloatsep}{10pt plus 2pt minus 10pt}
\begin{figure}
\centering
\includegraphics[width=\columnwidth]{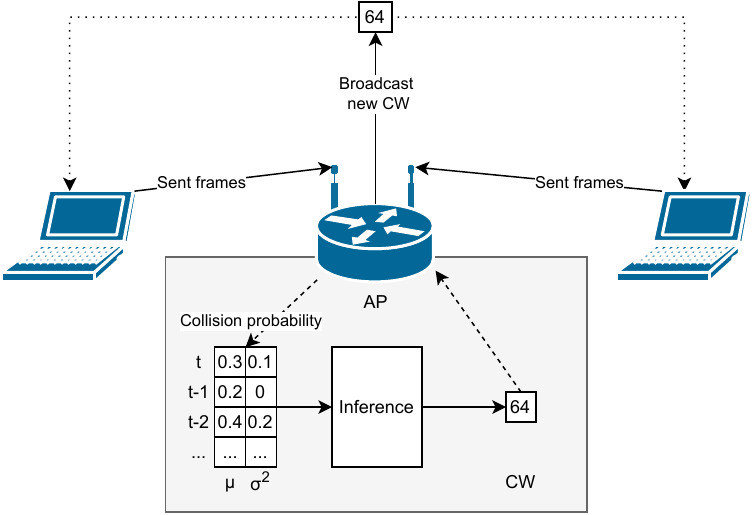}
\caption{The considered topology (stations transmit data to the AP) and process of CW update ($\mu$ denotes mean, $\sigma$ -- standard deviation).}
\label{fig:topology}
\end{figure}

The ns-3 simulations used the topology of Fig.~\ref{fig:topology} and the following settings: error-free radio channels, IEEE 802.11ax, the highest modulation and coding scheme (1024-QAM with a 	5/6 coding rate), single-user transmissions, a 20 MHz channel, frame aggregation disabled\footnote{Frame aggregation was disabled to speed up the experiments at the cost of throughput. This does not qualitatively affect the network behavior because if frame aggregation was enabled, the improvement would have been proportional to the gain in throughput.}, and constant bit-rate UDP uplink traffic to a single AP with 1500~B packets and equal offered load calibrated to saturate the network. 
Also, we assumed (\textit{i}) perfect and immediate transfer of state information to the agent (i.e., the current values of $N_{t}$ and $N_{r}$ are known at the AP) as well as (\textit{ii}) the immediate setting of $CW$ at each station. %
{In practice, relaxing the former assumption would require an overhead of around 100-200~B/s sent from the stations to the AP, while relaxing the latter assumption would require dissemination of CW values by the AP through periodic beacon frames leading to slower convergence.} 
In summary, the idealized simulation settings allow for assessing the base performance of CCOD before moving to more realistic topologies.

\begin{table}[t]
\centering
 \caption{CCOD's DRL settings}
 \begin{tabular}{@{}cc@{}} 
 \toprule
 Parameter & Value \\
 \midrule
 Interaction period & 10 ms \\
 History length $h$ & 300 \\
 DQN's learning rate & \num{4e-4} \\
 DDPG's actor learning rate & \num{4e-4} \\ 
 DDPG's critic learning rate & \num{4e-3} \\
 Batch size & 32 \\
 Reward discount $\gamma$ & 0.7 \\
 Replay buffer $B$ size & 18,000 \\
 Soft update coefficient $\tau$ & \num{4e-3} \\
 \bottomrule
 \end{tabular}

\label{table:drl-settings}
\end{table}

The DRL algorithms were run with the parameters in Table~\ref{table:drl-settings}, which were determined empirically through a lengthy simulation campaign to provide good performance for both algorithms (their universality is left for further study).
The hyperparameters were determined using a random grid search followed by Bayesian optimization. 
The neural network architecture was the same for both algorithms: one recurrent long short-term memory layer followed by two dense layers resulting in a $8\times128\times64$ configuration.
The size of the networks was determined in the same way as the other hyperparameters. Using a recurrent layer with a wide history window allowed the algorithms to take previous observations into account. The preprocessing window length was set to $\frac{h}{2}$ with a stride of $\frac{h}{4}$, where $h$ is the history length. 

Randomness was incorporated into both agent behavior and network simulation.
Each experiment was run for 15 rounds of 60-second simulations (the first 14 rounds constituted the learning phase, the last round -- the operational phase).
Each simulation consisted of 10 ms interaction periods, between which Algorithm~\ref{alg: cw_optim} was run.

\begin{figure}
\centering
\vspace{0.25cm}
\includegraphics[width=\myScaling\textwidth]{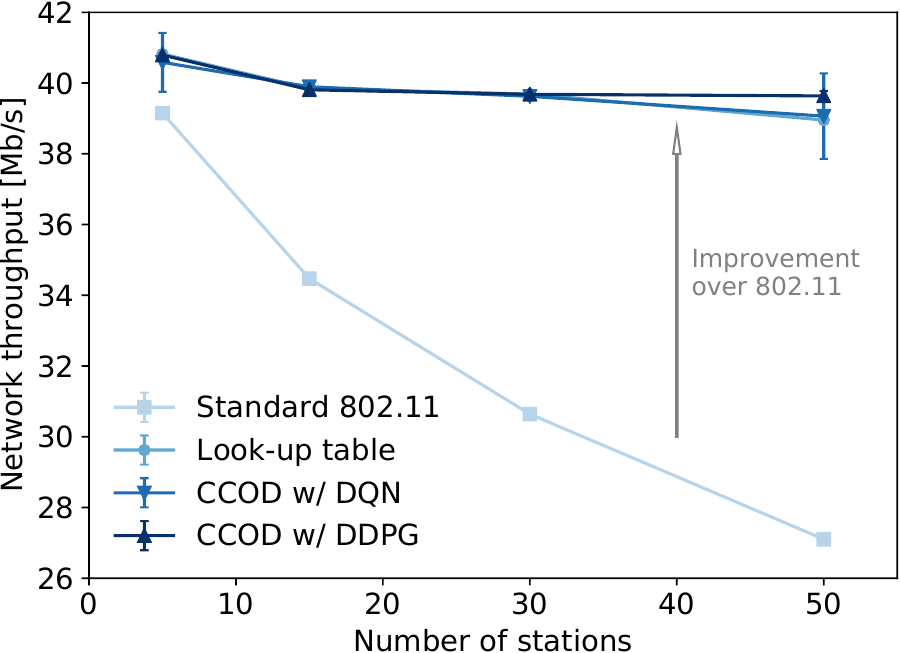}
\caption{Network throughput for the static topology.}
\label{fig:static}
\end{figure}
\section{Results}

CCOD was evaluated in two different topologies, for a static and dynamic number of stations, to assess various performance aspects. We used two baselines for comparison:
(a) the current operation of 802.11ax, denoted as \textit{standard 802.11}, in which $CW_{min} = 2^4-1$ and $CW_{max} = 2^{10}-1$, and 
(b) an idealized case of a \textit{look-up table} in which $CW_{min}=CW_{max}=CW$ and $CW \in \{2^x-1\ |\ x \in [4,10]\}$,  where $x$ depends on the number of stations currently in the network. 

The look-up table (a mapping between the number of stations and $CW$) was prepared \textit{a priori} by determining (with simulations) which CW values provide best network performance (for multiples of five stations).
The first baseline represents the current operation of 802.11ax, while the latter estimates the upper bound (under the assumption that only CW values being powers of 2 are available).
They were chosen to provide the best possible results in both topologies. %
Note that in all figures where the results are an average of multiple runs, the 95\% confidence intervals are shown or were too small for graphical representation.

Before proceeding to the network performance results, we report that one round of CCOD's training lasted for 24 minutes on an NVIDIA 1080ti GPU. However, much of the time was consumed by the network simulation and the overhead of the ns3-gym interface -- turning off the training and using a random agent resulted in 22 minutes of simulation. This means that the training itself caused only a 10\% slowdown. Since off-the-shelf APs are not equipped with GPUs, in real deployments the training would have to be done prior to installing the agent in the AP.

\subsection{Static Topology}
\begin{figure}
\centering
\includegraphics[width=\myScaling\textwidth]{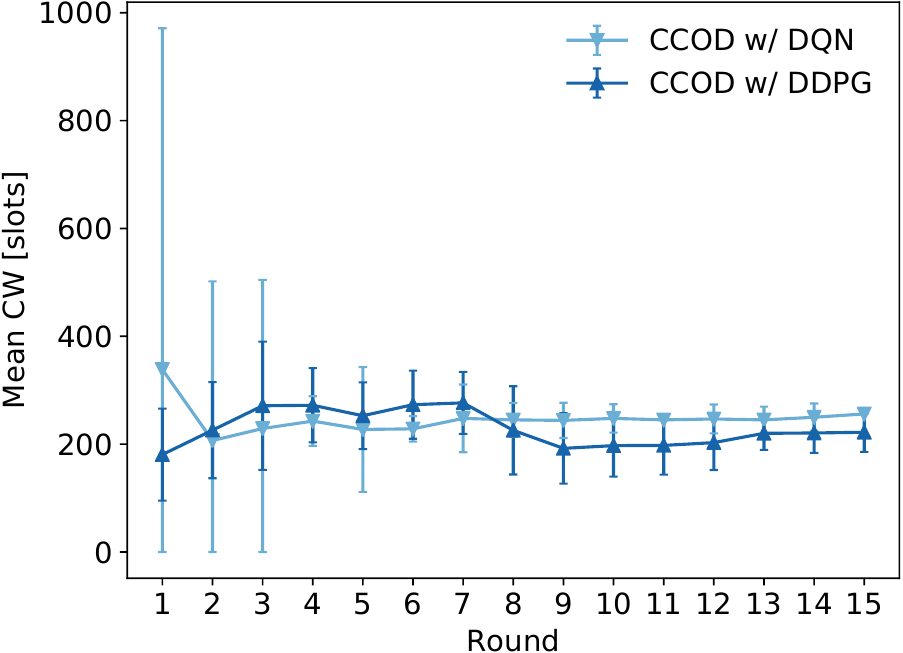}
\caption{Mean CW values selected in each round for 30 stations in the static topology.}
\label{fig:mean_cw}
\end{figure}
In the static topology, for each data point, there was a fixed number of stations connected to the AP throughout the simulation. In theory, a constant value of CW should be optimal in these conditions \cite{Gallo2018}. This scenario was designed to test whether CCOD's algorithms are able to recognize this value and what is the improvement over standard 802.11. For the look-up table approach, the CW values remained static throughout the experiment.

The results show that while 802.11 performance degrades for larger networks, CCOD with both DDPG and DQN can optimize the CW value in static network conditions (Fig.~\ref{fig:static}). 
The improvement over standard 802.11 ranges from $1.5\%$ (for $5$ stations) to $40\%$ (for 50 stations).
As anticipated, CCOD's operation reflects the performance of the look-up table approach (even slightly exceeding it for DDPG and 50 stations due to its ability to select any integer CW values and not just powers of 2). 

Fig.~\ref{fig:mean_cw} presents the mean CW value selected by both CCOD's algorithms in each round of simulating the static topology for 30 stations. 
Evidently the selected number of 14 rounds of the learning phase are enough to converge to stable CW values.

\subsection{Dynamic Topology}
\begin{figure}
\centering
\includegraphics[width=\myScaling\textwidth]{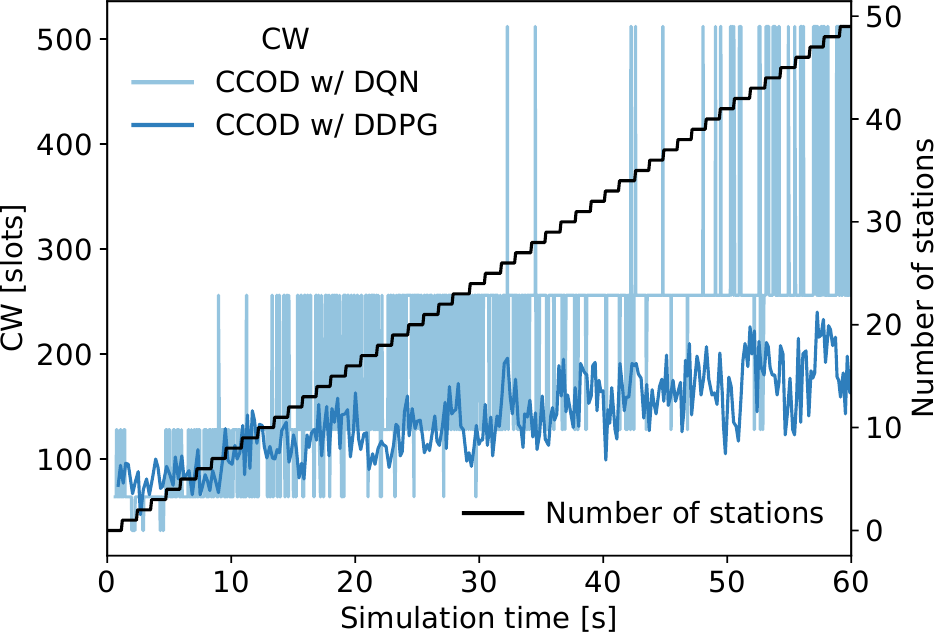}
\caption{CW values selected by CCOD for a given number of stations in the dynamic topology.}
\label{fig:cw_choice}
\end{figure}

\begin{figure}
\centering
\includegraphics[width=\myScaling\textwidth]{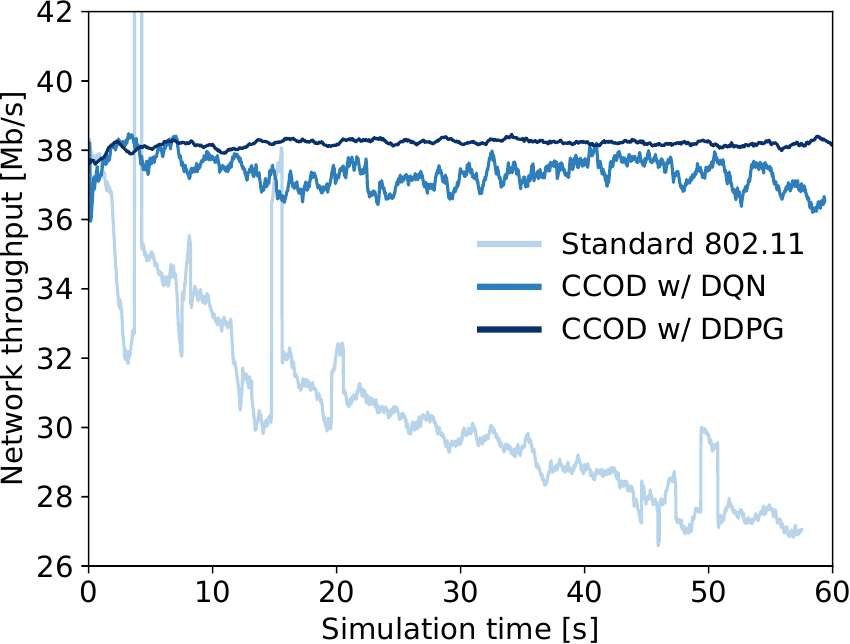}
\caption{Instantaneous network throughput for the dynamic topology. As the number of stations increases (from 5 to 50), the performance of standard 802.11 decreases, while CCOD is able to maintain a high throughput.}
\label{fig:reaction}
\end{figure}

In the dynamic topology, the number of transmitting stations steadily increased from 5 to 50 stations over the course of a single experiment, increasing the collision rate in the network.
This scenario was designed to test whether the algorithms are able to react to network changes. For the look-up table approach, the CW values were updated after every 5 stations joined the network.

\begin{figure}
\centering
\includegraphics[width=\myScaling\textwidth]{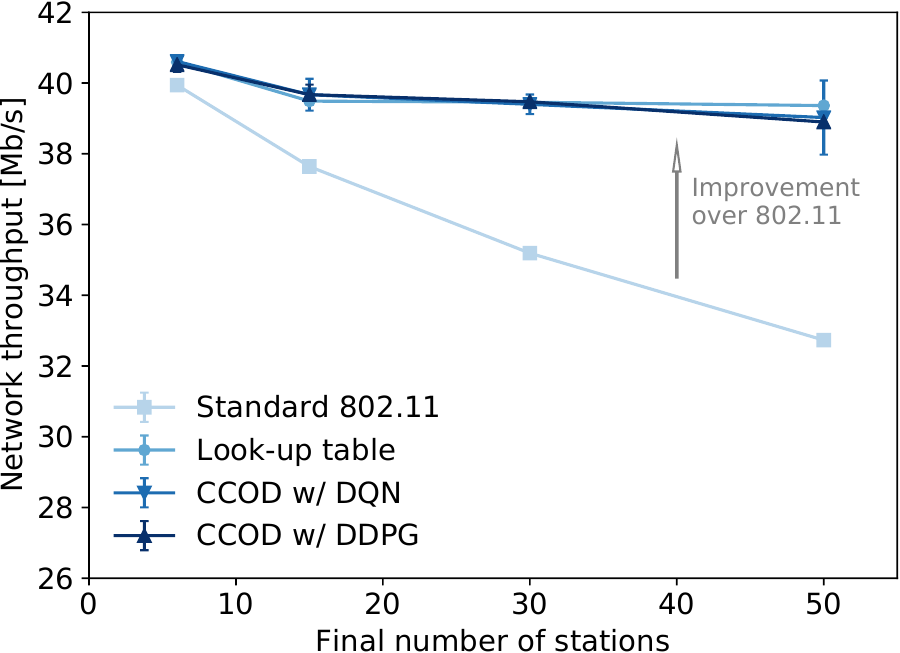}
\caption{Network throughput for the dynamic topology, where the number of stations steadily increased from 5 to the value on the X axis.}
\label{fig:dynamic}
\end{figure}

Fig.~\ref{fig:cw_choice} shows how the number of stations increased in a simulation run and how the CW values were updated accordingly. CCOD, with both algorithms, decides on increasing the CW value with the increasing number of stations. DQN strongly relies on oscillations between two (discrete) neighboring CW values as a way of increasing throughput. DDPG's continuous approach is able to follow the network behavior more closely, and (in this run) settled on  a lower final CW value.
The change in CW in each simulation run is reflected in the change of instantaneous throughput (Fig.~\ref{fig:reaction}). Standard 802.11 leads to a decrease of up to 28\% of the network throughput with the increasing number of stations. CCOD is able to maintain the efficiency on a similar level -- the decrease of throughput moving from 5 to 50 stations is only about 1\% for both DDPG and DQN.
Ultimately, the operation of both CCOD's algorithms in the dynamic topology lead to improved network performance (Fig.~\ref{fig:dynamic}), both exceeding standard 802.11 and matching the look-up table approach.

\section{Conclusions}

We have presented CCOD -- a method which leverages deep reinforcement learning principles to learn the correct CW settings for 802.11ax under varying network conditions using two trainable control algorithms: DQN and DDPG.
Our experiments have shown that DRL can be successfully applied to the problem of CW optimization: both algorithms offer efficiency close to optimal (with DDPG being only slightly better than DQN) even under rapidly changing topologies, while keeping the computational cost low (around 22 kflops, according to our estimations, excluding the one-time training cost). As a result of the learning process, we have obtained a trained agent which can be directly installed in an 802.11ax AP.

Additionally, our results confirm two well-known facts from 802.11 research: (1) performance can be improved by dynamically tuning CW based on network conditions \cite{Gallo2018} and (2) centralized schemes for CW tuning usually outperform decentralized ones \cite{Serrano2013}.
We conclude that the problem of CW optimization has provided the opportunity to showcase the features of DRL and its applications in 802.11-related research.
Future studies should focus on analyzing more realistic network conditions, with time-varying traffic loads, where we expect DRL to outperform any analytical model-based CW optimization methods which are based on simplifying  assumptions. Also worth investigating are other DRL algorithms as well as implementing a distributed version of CCOD. Another point of interest could be performing the optimization in regards to network latency to assess the relationship between delay and throughput in similar setups. 
Finally, deploying CCOD in real testbeds \cite{Sandholm2019}, possibly using FPGA \cite{Ding2019}, is worth investigating.

\section*{Acknowledgment}
This work was supported by the Polish Ministry of Science and Higher Education with the subvention funds of the Faculty of Computer Science, Electronics and Telecommunications of AGH University. This research was supported in part by PLGrid Infrastructure. 
The authors wish to thank Jakub Mojsiejuk for his remarks on an early draft of the paper.



\begin{thebibliography}{10}
\providecommand{\url}[1]{#1}
\csname url@samestyle\endcsname
\providecommand{\newblock}{\relax}
\providecommand{\bibinfo}[2]{#2}
\providecommand{\BIBentrySTDinterwordspacing}{\spaceskip=0pt\relax}
\providecommand{\BIBentryALTinterwordstretchfactor}{4}
\providecommand{\BIBentryALTinterwordspacing}{\spaceskip=\fontdimen2\font plus
\BIBentryALTinterwordstretchfactor\fontdimen3\font minus
  \fontdimen4\font\relax}
\providecommand{\BIBforeignlanguage}[2]{{%
\expandafter\ifx\csname l@#1\endcsname\relax
\typeout{** WARNING: IEEEtran.bst: No hyphenation pattern has been}%
\typeout{** loaded for the language `#1'. Using the pattern for}%
\typeout{** the default language instead.}%
\else
\language=\csname l@#1\endcsname
\fi
#2}}
\providecommand{\BIBdecl}{\relax}
\BIBdecl

\bibitem{Khorov2019}
E.~{Khorov}, A.~{Kiryanov}, A.~{Lyakhov}, and G.~{Bianchi}, ``{A Tutorial on
  IEEE 802.11ax High Efficiency WLANs},'' \emph{IEEE Communications Surveys \&
  Tutorials}, vol.~21, no.~1, pp. 197--216, 2019.

\bibitem{Bellalta2019}
B.~Bellalta and K.~Kosek-Szott, ``{AP-initiated multi-user transmissions in
  IEEE 802.11ax WLANs},'' \emph{Ad Hoc Networks}, vol.~85, pp. 145--159, 2019.

\bibitem{Gallo2018}
P.~{Gallo}, K.~{Kosek-Szott}, S.~{Szott}, and I.~{Tinnirello}, ``{CADWAN: A
  Control Architecture for Dense WiFi Access Networks},'' \emph{IEEE
  Communications Magazine}, vol.~56, no.~1, pp. 194--201, 2018.

\bibitem{Serrano2013}
P.~Serrano \emph{et~al.}, ``Control theoretic optimization of 802.11 {WLANs}:
  Implementation and experimental evaluation,'' \emph{Computer Networks},
  vol.~57, no.~1, pp. 258--272, 2013.

\bibitem{Karaca2017}
M.~{Karaca}, S.~{Bastani}, and B.~{Landfeldt}, ``{Modifying Backoff Freezing
  Mechanism to Optimize Dense IEEE 802.11 Networks},'' \emph{IEEE Tr. on
  Vehicular Technology}, vol.~66, no.~10, pp. 9470--9482, 2017.

\bibitem{Wilhelmi2019}
F.~{Wilhelmi}, S.~{Barrachina-Munoz}, B.~{Bellalta}, C.~{Cano}, A.~{Jonsson},
  and V.~{Ram}, ``{A Flexible Machine-Learning-Aware Architecture for Future
  WLANs},'' \emph{IEEE Comm. Mag.}, vol.~58, no.~3, pp. 25--31, 2020.

\bibitem{Sandholm2019}
T.~Sandholm, B.~Huberman, B.~Hamzeh, and S.~Clearwater, ``Learning to wait:
  Wi-fi contention control using load-based predictions,'' \emph{arXiv preprint
  arXiv:1912.06747}, 2019.

\bibitem{30years}
J.~{Wang}, C.~{Jiang}, H.~{Zhang}, Y.~{Ren}, K.~C. {Chen}, and L.~{Hanzo},
  ``{Thirty Years of Machine Learning: The Road to Pareto-Optimal Wireless
  Networks},'' \emph{IEEE Communications Surveys \& Tutorials}, vol.~22, no.~3,
  pp. 1472--1514, 2020.

\bibitem{Bianchi2000}
G.~{Bianchi}, ``Performance analysis of the {IEEE} 802.11 distributed
  coordination function,'' \emph{IEEE Journal on Selected Areas in
  Communications}, vol.~18, no.~3, pp. 535--547, 2000.

\bibitem{Zhang2019}
C.~{Zhang}, P.~{Patras}, and H.~{Haddadi}, ``Deep learning in mobile and
  wireless networking: A survey,'' \emph{IEEE Communications Surveys \&
  Tutorials}, vol.~21, no.~3, pp. 2224--2287, 2019.

\bibitem{Mnih2015}
V.~Mnih \emph{et~al.}, ``Human-level control through deep reinforcement
  learning,'' \emph{Nature}, vol. 518, no. 7540, p. 529, 2015.

\bibitem{DPG}
D.~Silver, ``Deterministic policy gradient algorithms,'' \emph{Proceedings of
  ICML'14}, vol.~32, pp. I--387--I--395, 2014.

\bibitem{Yao2019}
F.~{Yao} and L.~{Jia}, ``A collaborative multi-agent reinforcement learning
  anti-jamming algorithm in wireless networks,'' \emph{IEEE Wireless
  Communications Letters}, vol.~8, no.~4, pp. 1024--1027, 2019.

\bibitem{Yu2019}
Y.~{Yu}, T.~{Wang}, and S.~C. {Liew}, ``Deep-reinforcement learning multiple
  access for heterogeneous wireless networks,'' \emph{IEEE Journal on Selected
  Areas in Communications}, vol.~37, no.~6, pp. 1277--1290, 2019.

\bibitem{under-the-sea}
X.~Ye and L.~Fu, ``{Deep Reinforcement Learning Based MAC Protocol for
  Underwater Acoustic Networks},'' in \emph{Proc. of WUWNET}, 2019.

\bibitem{altam2020learn}
F.~AL-Tam, N.~Correia, and J.~Rodriguez, ``{Learn to Schedule (LEASCH): A Deep
  reinforcement learning approach for radio resource scheduling in the 5G MAC
  layer},'' 2020, arXiv:2003.11003.

\bibitem{Huang2020}
C.~{Huang}, R.~{Mo}, and C.~{Yuen}, ``{Reconfigurable Intelligent Surface
  oadisted Multiuser MISO Systems Exploiting Deep Reinforcement Learning},''
  \emph{IEEE Journal on Selected Areas in Communications}, vol.~38, no.~8, pp.
  1839--1850, 2020.

\bibitem{Ali2019}
R.~{Ali} \emph{et~al.}, ``{Deep Reinforcement Learning Paradigm for Performance
  Optimization of Channel Observation–Based MAC Protocols in Dense WLANs},''
  \emph{IEEE Access}, vol.~7, pp. 3500--3511, 2019.

\bibitem{gupta2018distributed}
O.~Gupta and R.~Raskar, ``Distributed learning of deep neural network over
  multiple agents,'' \emph{Journal of Network and Computer Applications}, vol.
  116, pp. 1 -- 8, 2018.

\bibitem{ns3gym}
P.~Gaw{\l}owicz and A.~Zubow, ``{ns-3 meets OpenAI Gym: The Playground for
  Machine Learning in Networking Research},'' in \emph{{ACM MSWiM}}, 2019.

\bibitem{Ding2019}
B.~{Ding}, J.~{Liu}, H.~{Wu}, and T.~{Wang}, ``{GPLM: An 802.11ac-Capable
  Low-MAC Architecture for FPGA-based SDR Systems},'' in \emph{2019 IEEE
  Wireless Communications and Networking Conference (WCNC)}, 2019.

\end{thebibliography}
\end{document}